\documentclass[aps, prb, longbibliography, reprint, noshowpacs, floatfix, superscriptaddress]{revtex4-2}

\usepackage{graphicx} 
\usepackage{bm} 
\usepackage{comment} 

\usepackage{hyperref}
\hypersetup{colorlinks,linkcolor=blue,urlcolor=blue,citecolor=blue}

\usepackage{xcolor}
\definecolor{lightgray}{gray}{0.6}

\newif\ifptitle
\newif\ifpnumber
\newcounter{para}
\newcommand\ptitle[1]{\par\refstepcounter{para}
{\ifpnumber{\noindent\textcolor{lightgray}{\textbf{\thepara}}\indent}\fi}
{\ifptitle{\textbf{[{#1}]}}\fi}}

\pnumbertrue

\newcommand{\BSCCO}{Bi$_2$Sr$_2$CaCu$_2$O$_{8+\delta}$}
\newcommand{\YBCO}{YBa$_2$Cu$_3$O$_{7-\delta}$}

\newcommand{\CaYBCOten}{Y$_{0.9}$Ca$_{0.1}$Ba$_2$Cu$_3$O$_{7-\delta}$}

\newcommand{\hphys}{Department of Physics, Harvard University, Cambridge, Massachusetts 02138, U.S.A.}
\newcommand{\heng}{School of Engineering \& Applied Sciences, Harvard University, Cambridge, Massachusetts 02138, U.S.A.}
\newcommand{\mpi}{Max-Planck Institut f$\ddot{u}$r Festk$\ddot{o}$rperforschung, D-70569 Stuttgart, Germany}
\newcommand{\uta}{Department of Physics, University of Texas at Austin, Austin, TX 78712, U.S.A.}
\newcommand{\bc}{Department of Physics, Boston College, Chestnut Hill, MA 02467, U.S.A.}
 
\begin{document}

\title{Alternate cleavage structure and electronic inhomogeneity in Ca-doped \texorpdfstring{\YBCO}{YBa2Cu3O7-x}}

\author{Larissa B. Little}
\affiliation{\heng}

\author{Jennifer Coulter}
\affiliation{\heng}

\author{Ruizhe Kang}
\affiliation{\heng}

\author{Ilija Zeljkovic}
\affiliation{\bc}

\author{Dennis Huang}
\affiliation{\hphys}
\affiliation{\mpi}

\author{Can-Li Song}
\affiliation{\hphys}

\author{Toshinao Loew}
\affiliation{\mpi}

\author{Han-Jong Chia}
\affiliation{\uta}

\author{Jason D. Hoffman}
\affiliation{\hphys}

\author{John T. Markert}
\affiliation{\uta}

\author{Bernhard Keimer}
\affiliation{\mpi}

\author{Boris Kozinsky}
\affiliation{\heng}

\author{Jennifer E. Hoffman}
\email{jhoffman@physics.harvard.edu}

\affiliation{\hphys}
\affiliation{\heng}

\date{\today}

\begin{abstract}
\YBCO\ (YBCO) has favorable macroscopic superconducting properties of $T_\mathrm{c}$ up to 93 K and $H_{c2}$ up to 150 T. However, its nanoscale electronic structure remains mysterious because bulk-like electronic properties are not preserved near the surface of cleaved samples for easy access by local or surface-sensitive probes. It has been hypothesized that Ca-doping at the Y site could induce an alternate cleavage plane that mitigates this issue. We use scanning tunneling microscopy (STM) to study both Ca-free and 10\% Ca-doped YBCO, and find that the Ca-doped samples do indeed cleave on an alternate plane, yielding a spatially-disordered partial (Y,Ca) surface. Our density functional theory calculations support the increased likelihood of this new cleavage plane in Ca-doped YBCO. On this surface, we image a superconducting gap with average value 24 $\pm$ 3~meV and characteristic length scale 1--2 nm, similar to Bi-based high-$T_\mathrm{c}$ cuprates, but the first map of gap inhomogeneity in the YBCO family. 
 \end{abstract}

\maketitle

\section{Introduction}
\ptitle{.}
High-$T_\mathrm{c}$ cuprates exhibit chemical, structural, and electronic disorders, which contribute to their rich phase diagram. Scanning tunneling microscopy and spectroscopy (STM and STS) are well-suited to study these atomic-scale properties across a range of temperatures and magnetic fields \cite{Fischer_RevModPhys_2007, Zeljkovic_PCCP_2013}. For example, STM and STS studies on cuprates have been used to quantify inhomogeneity in the spectral gap \cite{Cren_PRL_2000, Howald_PRB_2001, Lang_Nat_2002, McElroy_PRL_2005, Boyer_NatPhys_2007, Pushp_Sci_2009}, reveal ``checkerboard'' and other charge orders \cite{Howald_PNAS_2003, wang_real-space_2019}, and provide evidence for pair density wave order in some regions of the phase diagram \cite{HamidianNature2016, AgterbergARCMP2020}. However, only a subset of cuprates are suitable for STM study. Bulk crystals must: (a) cleave easily to reveal terraces of at least 50--100 nm, and (b) retain bulk-like superconductivity at the cleaved surface. Some cuprates such as La$_{2-x}$Sr${_x}$CuO$_4$ (LSCO) do not cleave, but rather fracture onto an uneven surface. Other cuprates such as \YBCO\ (YBCO) cleave to reveal a flat surface whose electronic structure does not reproducibly represent a bulk superconductor. Thus, most STM work has focused on a limited number of cuprates that fulfill both requirements, particularly \BSCCO\ (BSCCO) and other Bi-based cuprates. The limited scope of cuprate STM studies may not be fully representative of the rich high-$T_c$ cuprate material family. 

\begin{figure*}[http]
    \begin{center}
        \includegraphics[]{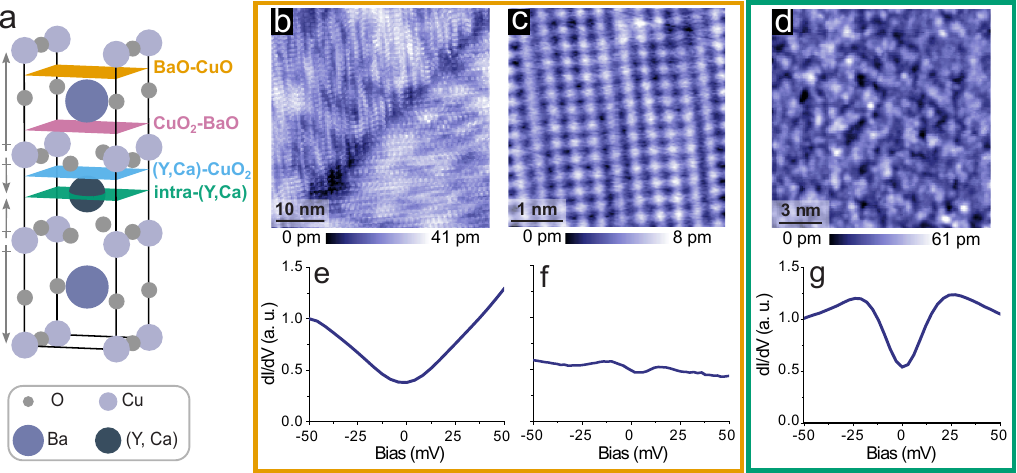}
        \caption{
            \label{fig:YBCOtopo}
            (a) YBCO unit cell with dipole moments shown as grey arrows. The four cleavage planes considered here are BaO-CuO (orange), CuO$_2$-BaO (pink), (Y-Ca)-CuO$_2$ (blue), and intra-(Y,Ca) (green). (b-g) STM topographs and corresponding $dI/dV$ spectra on various surfaces. (b,e) Optimally doped, Ca-free YBCO cleaved to reveal a CuO surface with a twin boundary, and no apparent superconducting gap.
            (c,f) Optimally doped, Ca-free YBCO cleaved to reveal a BaO square lattice with a gap that has a high zero bias conductance.
            (d,g) 10\% Ca-doped YBCO with hole doping $p = 0.12$ cleaved on the intra-(Y,Ca) plane to reveal a disordered surface morphology. Spectrum on the (Y,Ca) surface shows a superconducting gap.
            Experimental conditions: 
            (b,e) setpoint current $I_s=10$ pA and sample bias $V_s=-150$ mV;
            (c,f): $I_s=200$ pA, $V_s= -150$ mV;
            (d): $I_s=10$ pA, $V_s=-60$ mV; 
            (g): $I_s=50$ pA, $V_s=-60$ mV.
        }
    \end{center}
\end{figure*}

\ptitle{.}
YBCO is an appealing candidate for further STM work to contextualize existing studies on BSCCO. YBCO has high $T_\mathrm{c}$ (up to 93 K) and $H_{c2}$ (up to 150 T) in single crystal form \cite{Grissonnanche_NatComms_2014}. Due to its balance of anisotropy and flexibility, YBCO is also the primary cuprate used in commercial superconducting wire \cite{Molodyk_SciRep_2021}. However, cleaved YBCO does not reproducibly show bulk-like behavior at the surface. The Y plane and the CuO chain planes act as charge reservoirs for the superconducting CuO$_2$ planes [Fig.~\ref{fig:YBCOtopo}(a)]. When the undoped crystal is cleaved, it typically splits cleanly between the CuO and BaO planes. The internal dipole moment between the charge reservoir CuO and charge neutral BaO planes is therefore disrupted, leading to surfaces with a different electronic environment than the bulk crystal. See Appendix \ref{app:cleavage} for a comparison of BSCCO and YBCO cleavage.

\ptitle{.}
Given the electronic disruption upon cleaving YBCO, it is unsurprising that STM results have been varied. 
On the atomically-resolved CuO chain plane of cleaved YBCO, spectral gaps ranging from $\sim$18 to $\sim$30 meV were observed \cite{Edwards_PRL_1992, Maki_PhysC_2001, Maki_JPSJ_2001}, but several ambiguous gap-edge features were not consistently reproduced \cite{YBCO_gap_note}.
The complicated one-dimensional modulation and spectroscopic features observed on the CuO surface were initially attributed to a charge density wave \cite{Edwards_PRL_1994}, but later proposed to be proximity-induced superconductivity and quasiparticle scattering due to O vacancies in the CuO chains \cite{Derro_PRL_2002}. As-grown \cite{Maggio-Aprile_PRL_1995} and etched \cite{ShibataPhysicaC2003, Bruer_2014, Berthod_PRL_2017} YBCO surfaces have also been studied. These surfaces retain some bulk superconducting properties, as demonstrated by direct vortex imaging, but they lack the atomic resolution of cryogenically-cleaved YBCO that is needed to correlate the superconductivity with chemical or structural disorders.

\ptitle{.}
Doping YBCO with Ca has been proposed as a possible method for retaining bulk-like properties upon cleavage. Zabolotnyy \textit{et al.} performed angle-resolved photoemission spectroscopy (ARPES) measurements on the surface of cleaved Ca-free YBCO that revealed two components: (1) an overdoped metallic component with hole doping $p \sim 0.3$ regardless of the bulk hole concentration, presumed to come from the CuO$_{2}$ plane closest to the surface, and (2) a superconducting component presumed to originate from the CuO$_2$ planes further from the surface. However, in YBCO crystals with 15\% Ca substitution at the Y site, the overdoped surface component was suppressed. The authors hypothesized that the Ca substitution results in an alternate cleavage plane that preserves a bulk-like electronic environment for CuO$_2$ planes near the surface and thus allows better access to study the superconducting properties of the YBCO crystal \cite{Zabolotnyy_PRB_2007}.

\ptitle{.}
Here we study cleaved \YBCO\ and \CaYBCOten\ with STM and STS. We demonstrate experimentally and with support from density functional theory (DFT) calculations that Ca-doped YBCO cleaves through the (Y,Ca) crystal plane, revealing a surface that retains bulk-like superconductivity. On this (Y, Ca) surface, we observe a gap of $24\pm3$ meV with a characteristic length scale of 1--2 nm, similar to the spatial inhomogeneity of the superconducting gap reported on Bi-based cuprates. 

\section{Methods}
\subsection{Crystal preparation}
\ptitle{.}
We study single crystals of optimally-doped YBCO with $p\sim16\%$ and underdoped Ca-YBCO with $p\sim12\%$. The pure YBCO crystals were grown by a standard flux method in yttria-stabilized zirconia (YSZ) crucibles \cite{Edwards_PRL_1992}.
The Ca-YBCO crystals were synthesized by the solution-growth technique \cite{Zabolotnyy_PRB_2007} with $\sim$10\% Ca substitution, confirmed by energy-dispersive x-ray (EDX) measurements after the growth. Each Ca is expected to donate one hole to the bulk. Since each unit cell contains two CuO$_2$ planes per one Ca/Y plane, 10\% Ca substitution implies average 5\% hole doping per CuO$_2$ plane [Fig.~\ref{fig:YBCOtopo}(a)]. After synthesis, we annealed the crystals to deplete the oxygen and lower $p$ to a target value of 12\%. We first determined the annealing conditions necessary for Ca-free YBCO to have hole doping $p=7\%$, estimating the hole doping from the $c$-axis lattice parameter of the crystal \cite{Liang_PRB_2006}. Then, we annealed the 10\% Ca-doped YBCO under the same conditions to obtain a total hole doping of 12\%, assuming that the 5\% hole doping from the Ca and the 7\% hole doping from the oxygen vacancies add linearly. The optimally doped YBCO has $T_c=91$ K, while the Ca-YBCO has $T_c=31$ K, measured with magnetic susceptibility and shown in Appendix \ref{app:experiment}.

\subsection{STM and STS measurements}
\ptitle{.}
We cleave the YBCO and Ca-YBCO single crystals in ultra-high vacuum at cryogenic temperatures to expose a clean surface and immediately insert them into the STM head. We image the resulting surfaces at 6--7 K using PtIr tips cleaned by field emission on polycrystalline Au foil. All $dI/dV$ measurements are acquired using a standard lock-in amplifier technique with peak-to-peak amplitude 5 meV and frequency 1110 Hz. To measure the superconducting gap map, we acquire $dI/dV$ spectra on a square, densely-spaced pixel grid, and calculate the magnitude of the gap at each point by fitting a Gaussian curve to the more prominent gap-edge peak on the negative side of the spectrum. 

\subsection{DFT cleavage calculations}
\ptitle{.}
We use density functional theory (DFT) to investigate the cleavage of pure and Ca-doped YBCO. We compare the total energies of both pure and Ca-doped YBCO cleaved in each of the four configurations shown in Fig.~\ref{fig:YBCOtopo}(a): between the BaO and CuO planes [BaO-CuO], between the CuO$_2$ and BaO planes [CuO$_2$-BaO], between the (Y,Ca) and CuO$_2$ planes [(Y,Ca)-CuO$_2$], and finally through the (Y,Ca) atomic plane, with 50\% of the (Y,Ca) atoms on one of the resulting surfaces and 50\% on the other [intra-(Y,Ca)]. We use the intra-(Y,Ca) plane as an approximation of the disordered experimental behavior since true disorder is impractical to model with DFT.

\ptitle{.}
We use the VASP~\cite{vasp1,vasp2,vasp3} code with the projector-augmented wave method~\cite{vasp4PAW} and the r$^2$SCAN~\cite{r2SCAN} exchange-correlation functional.  As the challenges of studying correlated electron systems with DFT are well documented~\cite{r2SCAN}, we follow recent work that validated different methods of computing the structural properties of YBCO~\cite{r2SCAN_YBCO}. This work suggests that r$^2$SCAN used with an on-site $U$ parameter of 4 eV applied to the Cu $d$ orbitals and the D4 van der Waals correction is the most suitable method for describing the structural properties of YBCO~\cite{D4vdw}. For cleaved structure calculations, we found D4 to be numerically unstable, and therefore use the D3(BJ) vdW correction with damping parameters $a_1$ = 0.4948, $a_2$ = 5.7308, and $s_8$ = 0.7898, and a scaling factor $s_6$ = 1.0000~\cite{D3-BJ, r2SCAN-D4+D3Params}, which should produce a similar result~\cite{r2SCAN-D4+D3Params}. 

\ptitle{.}
We base our calculations on two initial crystal structures, YBa$_2$Cu$_3$O$_7$ and Y$_{0.8}$Ca$_{0.2}$Ba$_2$Cu$_3$O$_7$, taken from experimental results~\cite{fernandes_effect_1991, Jirak_PCS_1996}. Forces were converged using a plane-wave energy cutoff of 600 eV. To perform cleaved structure calculations, we use 2$\times$2$\times$2 supercells of the YBCO unit cell and a $\mathbf{k}$-mesh of 4$\times$4$\times$2. As DFT does not allow for fractional occupancies, we replace one of every four Y atoms in each of the Y atomic layers in the supercell with a Ca atom to generate a 25\% Ca-doped structure (Appendix \ref{app:DFT}). Lower Ca-doping is computationally impractical for cleavage calculations, given the large unit cell size.

\ptitle{.}
To calculate the surface energy of different cleavage planes, we follow standard methods \cite{Lazar_PRB_2008}. The unit cell structure was split, introducing 0.5 nm of vacuum between layers at the selected surface, as we observed minimal changes to the structure beyond this distance. This yields a slab thickness of 12 atomic layers upon cleavage. Then a structural relaxation with fixed unit cell size was performed to find the new atomic positions and system energies of the cleaved structures. Energies are reported as `relative energy per atom,' calculated by first finding the total energy of the cleaved system, subtracting the energy of the corrolary bulk system, then dividing by the number of atoms in the calculation. A visualization of the intra-(Y,Ca) cleavage plane is shown in Appendix \ref{app:DFT}.

\section{Results and Discussion}
\subsection{STM on cleaved YBCO and Ca-YBCO}

\ptitle{.}
We cleave optimally-doped Ca-free YBCO, and image both the CuO chains and BaO square lattice with atomic resolution. In Fig.~\ref{fig:YBCOtopo}(b), we see one-dimensional CuO chains with modulations, consistent with previous reports \cite{Edwards_PRL_1992, Edwards_PRL_1994, Edwards_JVCT_1994, Pan_RSI_1999, Maki_JPSJ_2001, Maki_PhysC_2001, Derro_PRL_2002, Nishizaki_JLTP_2003}. Although the surface is atomically smooth, there are 40 pm corrugations at the twin boundary. Our $dI/dV$ spectroscopy on this surface, shown in Fig.~\ref{fig:YBCOtopo}(e), exhibits no gap in our measured bias range, in contrast to some prior work
showing a $\sim$25 meV gap \cite{Edwards_PRL_1992, Maki_PhysC_2001, Maki_JPSJ_2001}. In Fig.~\ref{fig:YBCOtopo}(c), our BaO surface topography shows the same lattice seen in prior work \cite{Maki_JPSJ_2001, Edwards_JVCT_1994}.
Our $dI/dV$ spectroscopy on this surface, shown in Fig.~\ref{fig:YBCOtopo}(f), exhibits a weak partial gap, which has not been previously reported on the BaO surface to our knowledge. Our observed BaO gap may be explained by the close proximity of the BaO plane to the superconducting CuO$_2$ planes, coupled with the lack of charge on the BaO surface. 

\ptitle{.}
The Ca-doped YBCO crystal cleaves to reveal a surface with a new morphology, shown in Fig.~\ref{fig:YBCOtopo}(d). This surface is flat but disordered, with no step edges or valleys appearing over several hundred nanometer areas, as shown in Appendix \ref{app:experiment}. We do not achieve atomic resolution on this surface. Spectroscopy shows a symmetric gap in the density of states of 24 $\pm$ 3~meV around the Fermi level [Fig.~\ref{fig:YBCOtopo}(g)], consistent with the bulk gap \cite{Hufner_RPP_2008}. We suggest that Ca-YBCO crystals favor the intra-(Y,Ca) cleavage plane; i.e.\ they cleave through the (Y,Ca) atomic layer between adjacent CuO$_2$ layers, leaving a fraction of the (Y,Ca) atoms on either side. This partial (Y,Ca) occupancy is consistent with the disordered surface seen in Fig.~\ref{fig:YBCOtopo}(d). The CuO$_2$ plane alone is unlikely to explain the disordered surface, as atomically-resolved CuO$_2$ surfaces have been imaged in BSCCO \cite{Misra_PRL_2002, sugita_atomic_2000}. 

\subsection{DFT cleavage energy}
\begin{figure}
    \begin{center}
    \includegraphics[width=\columnwidth]{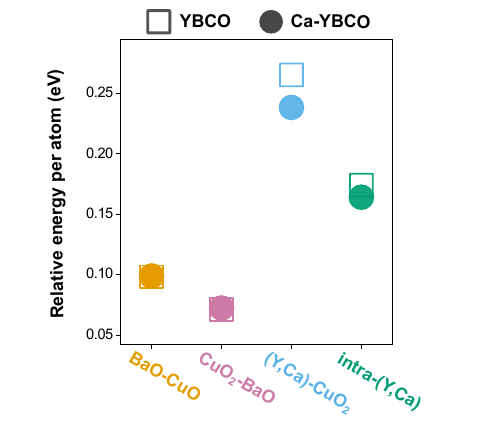}
    \caption{
        \label{fig:Surface_Energies}
        DFT-calculated surface energy of both YBCO and 25\% Ca-doped YBCO for each of the four possible cleavage planes. In line with experimental findings, Ca-doping increases the probability of cleaving to expose (Y,Ca) atoms by lowering the energy of both the (Y,Ca)-CuO$_2$ and the intra-(Y,Ca) planes, without significantly affecting the BaO-CuO or CuO$_2$-BaO plane energies. In contrast to experimental findings, the calculated overall surface energy of the CuO$_2$-BaO plane is the lowest of all plans for both YBCO and Ca-doped YBCO. See Appendix \ref{app:DFT}, Table~\ref{tab:dft_energies} for numerical values.
    }
    \end{center}
\end{figure}

\ptitle{.}
We use DFT to compare the energies of the four cleavage planes in Ca-doped and Ca-free YBCO, as shown in Fig.~\ref{fig:Surface_Energies}. We find that Ca-doping YBCO  decreases the energy cost of cleaving on the intra-(Y,Ca) or the (Y,Ca)-CuO$_2$ planes while minimally changing the cost of cleaving on the BaO-CuO or CuO$_2$-BaO planes. This trend supports that Ca-doped YBCO is more likely than Ca-free YBCO to cleave to expose the (Y,Ca) atoms, in line with our experimental results. Additionally, the calculations show that leaving 50\% of the (Y,Ca) atoms on each surface lowers the cleavage energy compared to cleaving between the Y and CuO$_2$ planes, again in line with our experimental results. This trend holds true across a range of oxygen stoichiometries (Appendix \ref{app:DFT}).

\ptitle{.}
Though the relative energies between Ca-doped and Ca-free YBCO are in line with our experimental observations, the relative energies between different cleavage plans of a single YBCO compound exhibit discrepancies. In our calculations, the CuO$_2$-BaO plane has the lowest surface energy of the four possible cleavage planes, indicating that it should be the most favorable. This calculated result is inconsistent with experiments on Ca-free YBCO, where the CuO-BaO plane is the most commonly observed. There are several possible explanations for this discrepancy. Some long-range ordering that is not captured in our moderately-sized unit cells could play an important role in inter-layer bond strength. Experiments have seen charge density waves phases in YBCO with a period of four unit cells \cite{Proust_remarkable_2019}, while prior DFT work on YBCO indicates that there are a number of stripe phases very close in energy with periodicities of several unit cells \cite{Zhang_PNAS_2020}. Furthermore, experiments indicate a disordered surface with partial (Y,Ca) occupancy, while disorder cannot be accurately captured in DFT. Finally, and perhaps most importantly, we know that YBCO is a strongly correlated material and the shortcomings of DFT in handling electronic interactions may not adequately capture inter-layer bond strength. 

\subsection{Superconductivity on (Y,Ca) surface}

\begin{figure}
    \begin{center}
    \includegraphics[width=\columnwidth]{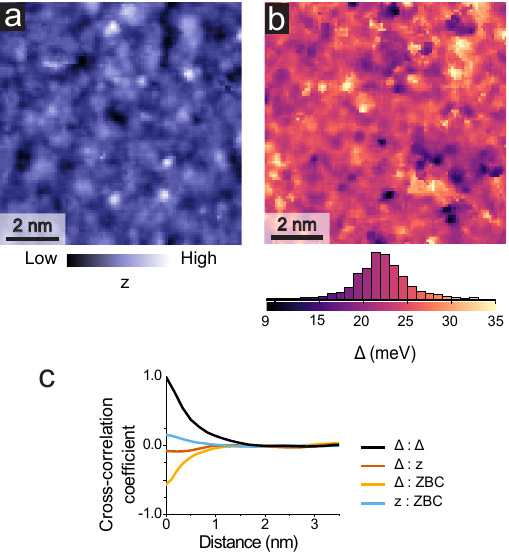}
    \caption{
        \label{fig:YBCOgapmap}
        (a) Topograph and (b) spectral gap map of \CaYBCOten\ acquired simultaneously over a 10 nm square region ($I_s=50$ pA, $V_s=-60$ mV).
        (c) Correlation coefficients for gap $\Delta(r)$, zero bias conductance (ZBC), and topographic height $z$. The auto-correlation coefficient of $\Delta(r)$ shows a characteristic gap length scale of 1--2 nm. The cross-correlation coefficient between gap and topography is low. 
    }
    \end{center}
\end{figure}

\ptitle{.}
We explore the nanoscale electronic structure by acquiring a simultaneous topography and gap map, in Fig.~\ref{fig:YBCOgapmap}.
The gap $\Delta(r)$ has an average value of $24 \pm 3$ meV, with extremal values from 9 to 35 meV over the 10 nm field of view. This relative variation is comparable to that seen throughout the full doping range in Bi-based cuprates \cite{McElroy_PRL_2005, Pushp_Sci_2009, he_fermi_2014}.
The auto-correlation coefficient of the gap map in Fig.~\ref{fig:YBCOgapmap}(b) shows a length scale of 1--2 nm [Fig.~\ref{fig:YBCOgapmap}(c)], also comparable to the length scale observed in Bi-based cuprates \cite{McElroy_PRL_2005,Pushp_Sci_2009, he_fermi_2014}. We find negligible correlation between the topographic height [$z(r)$], spectral gap [$\Delta(r)$)], and zero bias conductance [ZBC$(r)$] features [Fig.~\ref{fig:YBCOgapmap}(c)], suggesting that the spatial distribution of the spectral gap and ZBC is independent of surface disorder. Together with ARPES experiments on cleaved 15\% Ca-doped YBCO that detected spectral weight attributed to Cooper pairing \cite{Zabolotnyy_PRB_2007}, our data suggest that the gap we observe on the new surface can be identified as a superconducting state with spatially inhomogeneous properties similar to that of other cuprates. Spectral gap inhomogeneity seems to be omnipresent in Bi-based cuprates, but gap inhomogeneity was not previously reported in YBCO \cite{Maki_PhysC_2001}.

\ptitle{.}
Previous STM studies have imaged a disordered vortex lattice in near-optimally-doped BSCCO \cite{Pan_PRL_2000, Hoffman_Sci_2002} and YBCO \cite{Berthod_PRL_2017, Maki_JPSJ_2001, Maggio-Aprile_PRL_1995}. Vortices in these earlier maps appear as regions of higher ZBC, and lower $dI/dV$ at energies close to the gap edge. In ten tip-sample approaches in magnetic field on two different zero-field-cooled Ca-YBCO samples, we have not observed a vortex lattice [Fig.~\ref{fig:Bfield_topos}]. We suggest the formation of a vortex liquid, similar to that observed in other underdoped cuprates, including closely related compound YBa${_2}$Cu${_4}$O${_8}$ \cite{hsu_anomalous_2021}. 

\ptitle{.}
Though we do not observe static vortices in Ca-doped YBCO, we observe that the gap fills slightly with increasing $B$-field [Fig.~\ref{fig:Bfield_spectra}]. To the best of our knowledge, $H_{c2}$ of 10\% Ca-doped YBCO has not been reported, so we have no good reference point for the expected behavior of the superconducting gap up to 9 T. In Ca-free YBCO, $T_\mathrm{c}$ is more sensitive to magnetic field at hole doping around $p=0.12$ ($H_{c2} = 25$ T) compared to optimal hole doping $p = 0.16$ ($H_{c2} > 150$ T) \cite{Cyr-Choinière_PRB_2018, Grissonnanche_NatComms_2014}. However, this $H_{c2}$ dip at $p=0.12$ is attributed to competition from the incipient charge density wave, which has not been reported in 10\% Ca-doped YBCO and which we do not observe in our sample [Fig.~\ref{fig:YBCOtopo}(d)]. More data on the high-$B$ suppression of $T_\mathrm{c}$ in bulk Ca-doped YBCO are necessary to provide a point of comparison for our STM measurements of Ca-doped YBCO up to $B=9$ T.

\section{Conclusion}

\ptitle{.}
In summary, we study the effect of Ca-doping on the preferred cleavage plane of YBCO and characterize the resulting superconducting partial (Y,Ca) surface with STM. In alignment with prior work, our measurements of cleaved Ca-free, optimally-doped YBCO crystals show an electronic environment that is not representative of bulk and thus not compatible with the use of surface-sensitive techniques to study the nanoscale superconducting properties. However, when YBCO is doped with 10\% Ca, crystals cleave to expose a partial (Y,Ca) surface that is consistent with bulk-like superconductivity. DFT calculations support the increased probability of this intra-(Y,Ca) cleavage plane in Ca-doped samples, relative to the commonly observed BaO-CuO cleavage plane in Ca-free samples, though the calculated relative energies of the different cleavage planes within a single sample do not align with experimental results. We use STM to study the superconductivity on the partial (Y,Ca) surface and find a superconducting gap of 24 $\pm$ 3~meV with length scale 1--2 nm, similar to that seen in Bi-based cuprates. Ours is the first report of spatial inhomogeneity of the spectral gap on YBCO compounds. Areas of opportunity for Ca-doped YBCO include cleaving at a lower temperature to achieve a more ordered surface, and imaging a Ca-doped compound with optimal hole concentration $p=0.16$ to search for a vortex lattice. Our work opens the door for further study of the superconductivity in Ca-doped YBCO via surface-sensitive techniques like STM, and demonstrates the use of doping to modify the cleavage plane of crystals and reveal new surfaces of interest.

\begin{acknowledgments}
We thank E.~W.~Hudson and K.~Bystrom for helpful conversations. Pure YBCO crystals were grown and characterized by H.-J.~C. and J.~T.~M., supported by NSF DMR-0605828 and the Robert A. Welch Foundation Grant F-1191.
STM experiments were performed by I.~Z., D.~H., and C.-L.~S., with preliminary data and figures appearing in I.~Z.'s PhD thesis \cite{Zeljkovic_thesis_2013}, supported by Air Force Office of Scientific Research (AFOSR) grant FA9550-05-1-0371. Exploratory DFT calculations, crystal structure determination, and writing by L.~B.~L. were supported by the AFOSR under MURI grant FA9550-21-1-0429, and National Defense Science and Engineering Graduate Fellowship (NDSEG) awarded by the Air Force. Production quality cleaved cell calculations by J.~C. were supported by DOE grant DE-SC0020128. D.~H.\ acknowledges support from an NSERC PGS-D fellowship. C.~L.~S.\ acknowledges support from the Golub Fellowship at Harvard University.
\end{acknowledgments}

\clearpage
\appendix

\section{YBCO vs.\ BSCCO}
\label{app:cleavage}

\begin{figure}[htbp!]
\begin{center}
\includegraphics[width=\columnwidth]{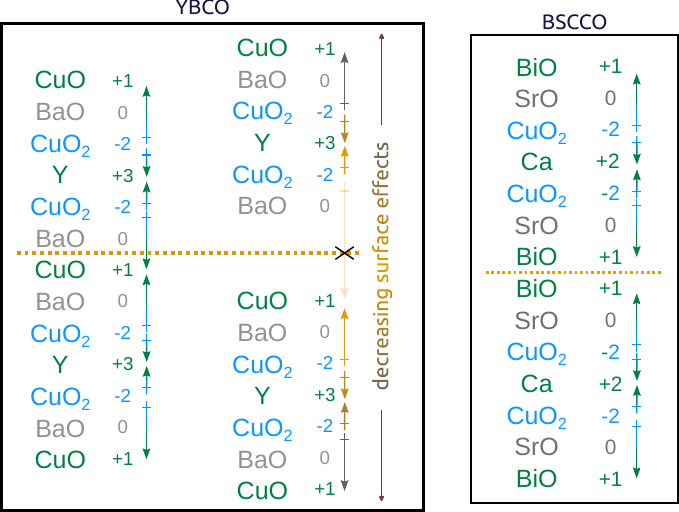}
\caption{\label{fig:dipole}Schematic of the dipole moments and cleavage planes in YBCO and BSCCO.}
\end{center}
\end{figure}

YBCO commonly cleaves between the BaO square lattice and CuO chain planes, which intersects an internal dipole moment. In contrast, BSCCO commonly cleaves between two van-der-Waals-bonded BiO layers that have mirror symmetry. Thus BSCCO cleavage does not interrupt any dipole moments and allows the surface to retain bulk-like electronic properties. The comparison between YBCO and BSCCO cleavage is shown in Fig.~\ref{fig:dipole}.

\section{Experimental characterization of C\MakeLowercase{a}-YBCO}
\label{app:experiment}

\begin{figure}[htbp!]
\begin{center}
\includegraphics[width=\columnwidth]{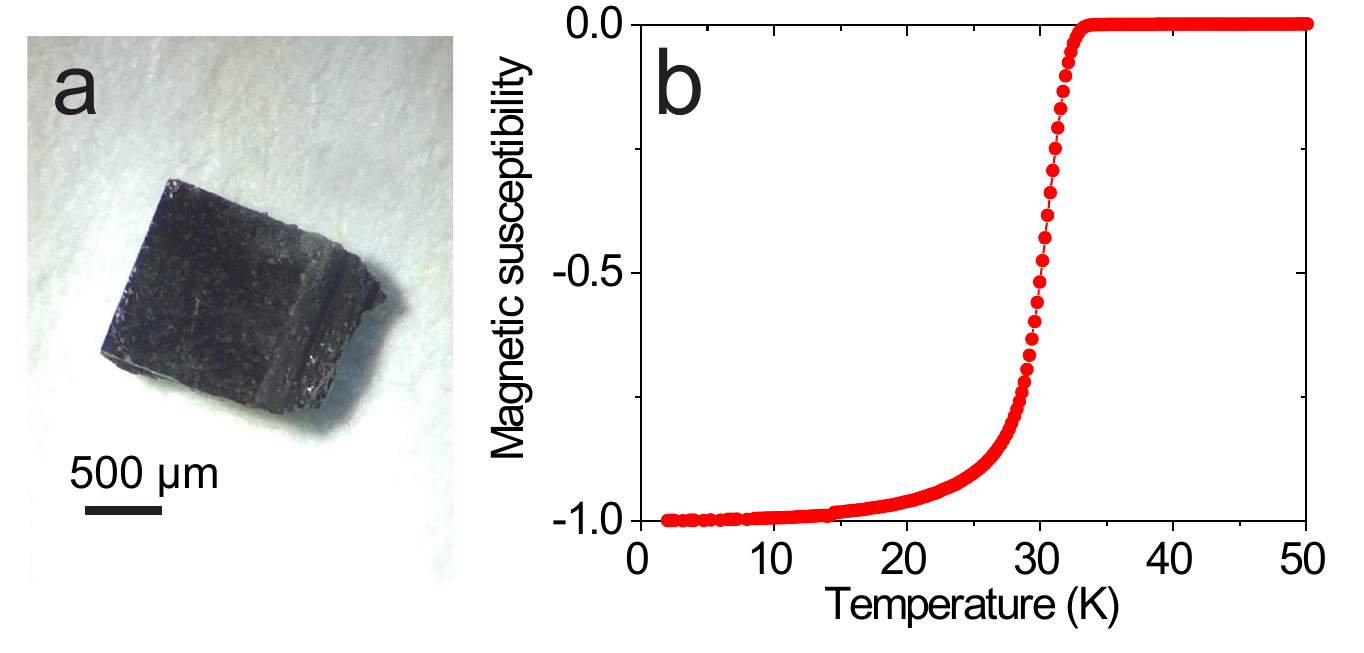}
\caption{\label{fig:susceptibility} (a) Photograph of one of the Ca-YBCO single crystals from the same batch used in this study. (b) Magnetic susceptibility measurements of Ca-YBCO samples with $T_\mathrm{c} \sim $ 31 K determined as the peak in the first derivative of the susceptibility vs.\ temperature trace.
}
\end{center}
\end{figure}

\begin{figure*}[http]
\begin{center}
\includegraphics[width=1.8\columnwidth]{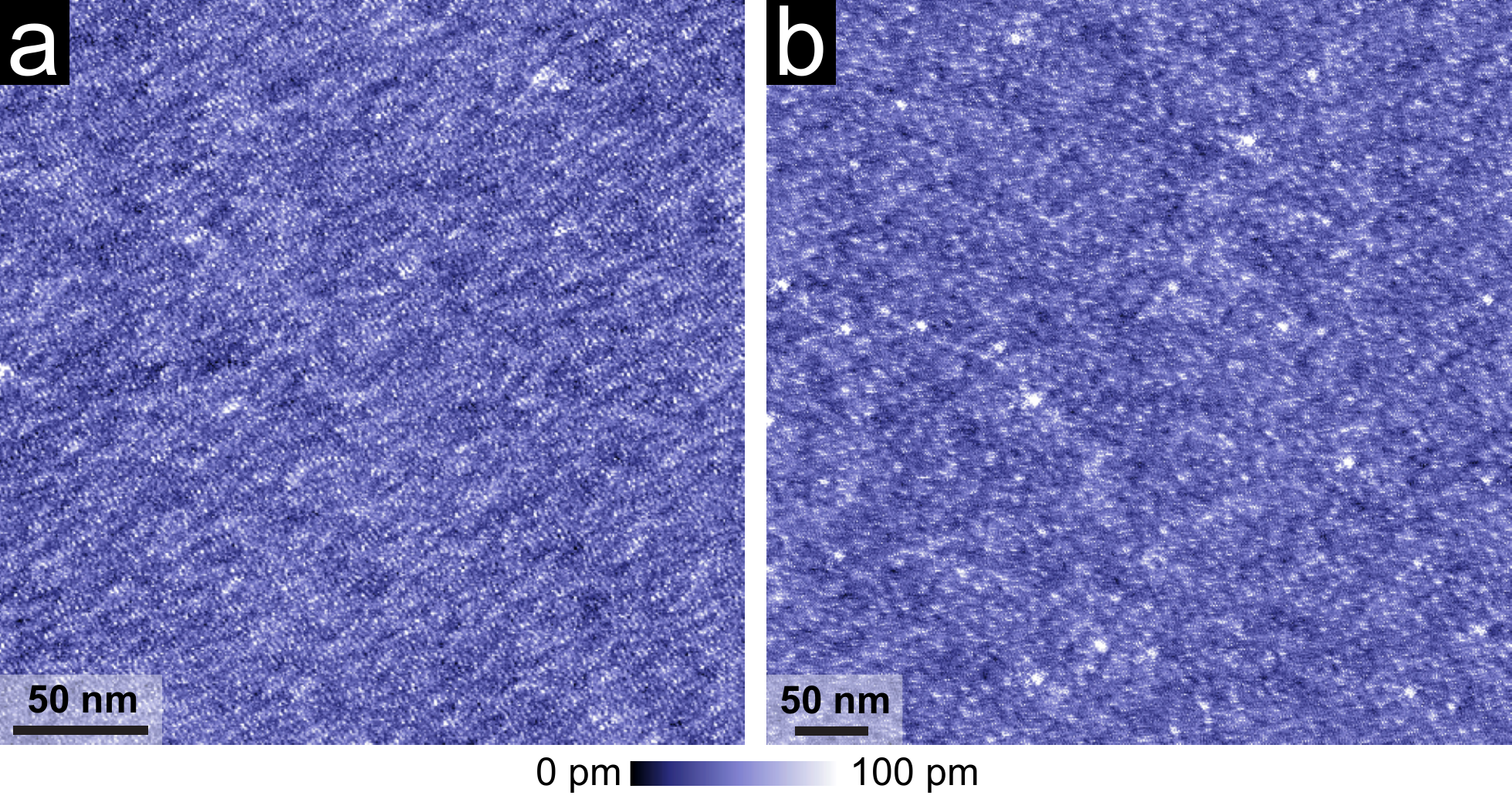}
\caption{\label{fig:large_topo} STM topographs of two different cryogenic cleaves of Ca-YBCO showing (a) $\sim$300 nm, and (b) $\sim$500 nm square regions of the sample.  Setup condition in (a): $I_s=10$ pA and $V_s=-500$ mV; (b) $I_s=10$ pA, $V_s=-150$ mV.
}
\end{center}
\end{figure*}

We performed STM experiments on two different cleaved \CaYBCOten\ crystals from a single batch. Magnetic susceptibility measurements on a third crystal from the same batch showed a T$_c = 31$ K (Fig.~\ref{fig:susceptibility}). STM topographies over several hundred nanometer areas show flat but disordered surfaces, with no steps, valleys, or reconstructions on either of the two cryogenic cleaves (Fig.~\ref{fig:large_topo}).

\begin{figure*}[!http]
\begin{center}
\includegraphics[width=1.8\columnwidth]{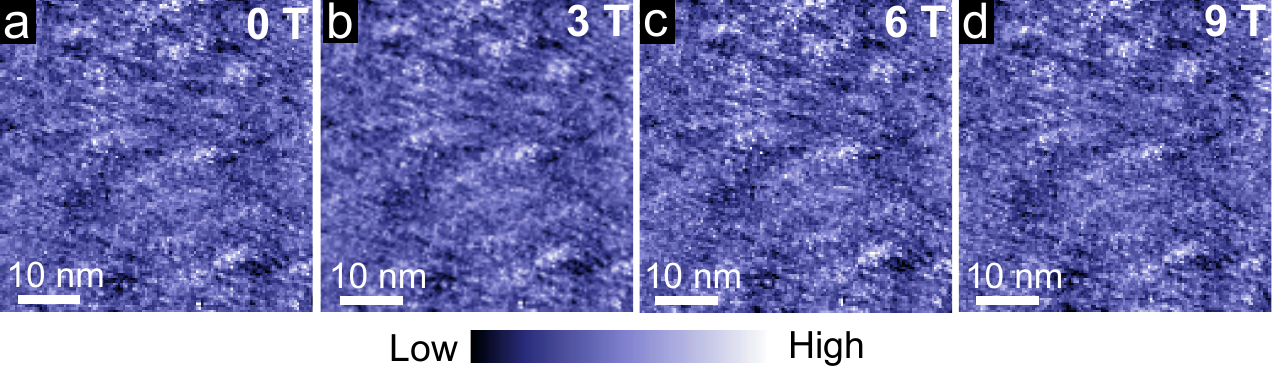}
\caption{\label{fig:Bfield_topos} Topographs over the same region of the sample acquired at (a-d) 0, 3, 6, and 9 T magnetic fields respectively.
}
\end{center}
\end{figure*}

We track a single region of the cleaved Ca-YBCO surface as $B$ is increased to 9 T in Fig.~\ref{fig:Bfield_topos}. 
Average spectra from this region, displayed in Fig.~\ref{fig:Bfield_spectra}, show that the gap fills only slightly with increasing $B$.

\section{DFT Calculations}
\label{app:DFT}

\begin{figure*}[http!]
\begin{minipage}[b]{0.45\linewidth}
\centering
\includegraphics[width=\columnwidth]{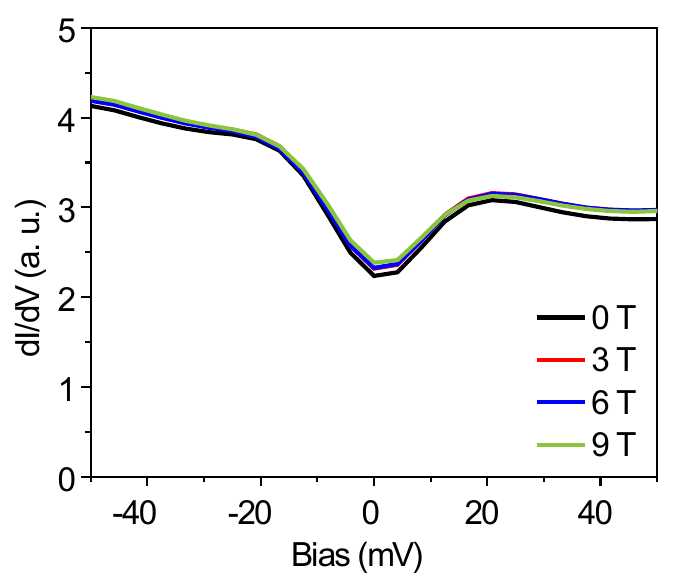}
\caption{\label{fig:Bfield_spectra} Average $dI/dV$ spectra obtained over the region shown in Fig.~\ref{fig:Bfield_topos}.  The Fermi-level gap fills only slightly with application of magnetic field up to 9 T.
}\end{minipage}
\hspace{0.5cm}
\begin{minipage}[b]{0.45\linewidth}
\centering
\includegraphics[width=\columnwidth]{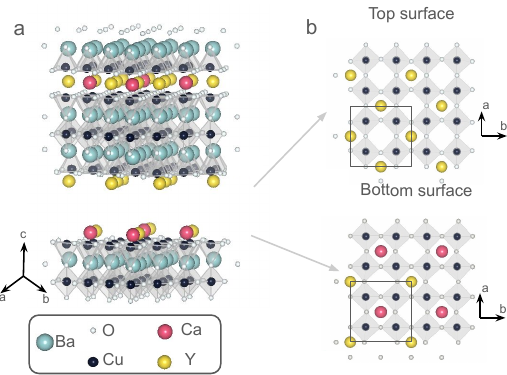}
\caption{\label{fig:intra_Y_cleave} (a) Side view of the intra-(Y, Ca) cleavage planes and (b) top view of the two resulting partial (Y,Ca) surfaces with black square showing the DFT supercell.}
\end{minipage}
\end{figure*}

In our intra-(Y,Ca) cleavage calculations, the partial (Y,Ca) surfaces have ordering not present in our experimental results: all of the Ca-atoms end up on one of the two surfaces, and there is a $2\times2$ `stripe-like' supercell that we do not observe in our STM images on the partial (Y,Ca) surfaces (Fig.~\ref{fig:intra_Y_cleave}).

To study the effect of oxygen doping on the effect of Ca-doping, we calculate cleavage energies for Y$_{0.75}$Ca$_{0.25}$Ba$_2$Cu$_3$O$_6$ (Ca-YBCO6) in addition to the YBa$_2$Cu$_3$O$_7$ (YBCO7) and Y$_{0.75}$Ca$_{0.25}$Ba$_2$Cu$_3$O$_7$ (Ca-YBCO7) structures discussed in the main text. The bulk Ca-YBCO6 structure was taken from experiment \cite{parise_structure_1989} and differs from the Ca-YBCO7 structure in that the apical O is removed. In other words the `CuO' planes become Cu planes, while the CuO$_2$ planes remain unchanged. Results of the cleavage calculation show that the trends discussed in the main text hold for both Ca-YBCO7 and Ca-YBCO6 (Table~\ref{tab:dft_energies}).

\begin{center}
\begin{table}[hb]
\caption{\label{tab:dft_energies} Energy difference per atom (eV) for each structure and cleavage plane.}
\begin{ruledtabular}
\begin{tabular}{c c c c}
Cleavage plane & YBCO7 & Ca-YBCO7  &  Ca-YBCO6 \\
\hline
 BaO-CuO        & 0.098  & 0.099       & 0.114   \\ 
 CuO$_2$-BaO    & 0.071  & 0.072       & 0.112   \\
 (Y,Ca)-CuO$_2$ & 0.265  & 0.234       & 0.251   \\
 intra-(Y,Ca)   & 0.174  & 0.164       & 0.166   \\
\end{tabular}
\end{ruledtabular}
\end{table}
\end{center}

\newpage

\vspace*{2in} 
\bibliography{YBCO_refs} 

\end{document}